# Maximizing the scientific return of Roman and Rubin with a joint wide-sky observing strategy

**Submitting Author:**

Name: Federica B. Bianco

Affiliation: University of Delaware / Rubin Observatory

Email: fbianco@udel.edu

**List of contributing authors** (including affiliation and email): (in alphabetical order)

Robert Blum, Rubin Observatory, bob.blum@noirlab.edu
Andrew Connolly, U. Washington/Rubin Observatory, ajc@astro.washington.edu
Melissa Graham U. Washington/Rubin Observatory, mlg3k@uw.edu
Leanne Guy Rubin Observatory, leanne@lsst.org
Zeljko Ivezic U. Washington/Rubin Observatory, ivezic@uw.edu
Steve Ritz, UC Santa Cruz/Rubin Observatory, sritz@ucsc.edu
Michael A. Strauss, Princeton, strauss@astro.princeton.edu
Tony Tyson, UC Davis/Rubin Observatory, tyson@physics.ucdavis.edu

## Abstract:

This proposal presents the case for a single-band LSST-matched depth Roman Community Survey over the footprint of the Vera C. Rubin Observatory Legacy Survey of Space and Time (LSST) Wide-Fast-Deep to enhance the key science programs of both missions. We propose to observe the ~18K sq deg LSST Wide-Fast-Deep footprint in the *F146* filter to *mAB*~25; this will take approximately 5 months of Roman observing time. The combination of the multiwavelength nature of LSST and angular resolution of Roman would lead to enhanced scientific returns for both the Roman and LSST surveys. Galaxy deblending and crowded field photometry will be significantly improved. The extension of Rubin LSST six-band optical photometry to IR wavelengths would improve photometric redshift (photo-z) estimation, leading to improved cosmological parameter estimation, penetrate interstellar dust in the Galactic plane, improve differential chromatic refraction derived SEDs, maximize galaxy-star separation and minimize crowding confusion through improved angular resolution. Conversely, the LSST survey will provide a time-domain extension of the Roman survey on the shared footprint and 6-band optical photometry with sensitivity extending all the way to ultraviolet wavelengths.

The Vera C. Rubin Observatory will start an unprecedented survey of the night sky in 2025, the Legacy Survey of Space and Time (LSST [1]), which will continuously collect multi-filter optical images of the entire southern hemisphere sky (~18,000 square degrees) for 10 years. The LSST is designed to deliver a transformational dataset to probe dark matter and dark energy, Solar System science, a wide variety of transient and variable astrophysics, and Galactic science. However, its reach will extend to many other areas of astrophysics, such as galaxy evolution and morphology, AGN/quasar astrophysics, and stellar populations. The Rubin system is designed to simultaneously maximize field of view and depth, *i.e.* to maximize the volume of space-time that can be surveyed. The optical wavelength range (0.35-1.05 micron) and time-resolved nature of LSST, which will probe time scales from minutes to years, provides a highly complementary dataset to Roman's, opening opportunities to enhance the science of both missions.

While Rubin LSST is designed to be transformational on its own, the survey potential can be further enhanced by synergistic operations with other surveys, including joint catalog-level analyses, and joint pixel-level processing of the independent survey data sets. Coordinated observing with space facilities can augment the LSST data by probing wavelengths inaccessible from the ground in an atmosphere-free environment. Several white- and peer-review papers on this topic have been published and Rubin is considering this input in the optimization of the Rubin observing strategy [2]. These include synergy with Roman [3, 4], Euclid [5, 6], and Roman+Euclid [7, 8].

While a multi-epoch / multi-filter full-sky Roman survey would be unfeasible, **combining Roman and LSST data would enable a full-sky Roman legacy**. A key element of the LSST legacy is the full-sky nature of the survey (within the limitation of a ground-based search, which enables only the observation of one sky hemisphere). Just as the Sloan Digital Sky Survey has done for the northern hemisphere, the coverage of Rubin LSST will be key in transforming our understanding of the Universe, enabling a multitude of studies for decades, unplanned discoveries, and unexpected science results. The proposed survey, especially if accompanied by Roman-Rubin coordinated observing on a smaller footprint in all Roman filters (as already planned, and as suggested in many other proposals including Hirata *et al.*, Newman *et al.*, Street *et al.*, etc) would maximize the Roman-Rubin synergy. Bootstrapping on these narrow but deeper, multi-band Roman extragalactic and Galactic surveys, the proposed wide survey will enable inference, not just estimation of errors, on one order of magnitude more astrophysical objects than the small-footprint surveys alone, enabling, for example, exquisite tomographic cosmic shear measurements by improved photo-z and improved star-galaxy separation.

The characteristics that are most important in maximizing synergistic operations between Rubin and Roman are: **(1) the extension of the wavelength coverage from optical to IR wavelengths which will benefit both surveys; (2) the spatial resolution and astrometric accuracy offered by the space-based, atmosphere-free platform that Roman operates in, which will enhance spatial characterization in Rubin images (3) the multi-filter, temporally rich information that Rubin provides, which complements the Roman sky survey.** Whereas other authors are advocating for additional surveys that would enable and enhance specific science goals, here we argue for a **simple, cadence-flexible, and affordable in terms of observing time, Core Community Survey:** *the coverage of the full LSST footprint with Roman in a single filter imaging campaign with LSST-matched depth.*

The current plan for the LSST observing strategy, including footprint details and per-filter depth, is described in PSTN-055 [9] (Figure 1). The LSST survey comprises an optimized mix of Wide-Fast-Deep (WFD) and Deep Drilling Field (DDF) observations, along with several mini and micro surveys. The WFD area is the core element of the LSST survey that is relevant here, in which the proposed observing strategy results in 18,620 sq. degrees of sky observed at least 750 times (summed over the six filters), with a median of 829 observations and a coadded point-source 5-σ depth of $r$~26.7 (as derived from current survey simulations and laboratory-measured system throughputs [10]). **The footprint of the proposed Roman survey should be designed to match the LSST WFD footprint. The cadence details are of secondary importance, although maximizing the temporal overlap of the two surveys could open up new possibilities. The principal remaining choices then are the optimal filter and the optimal depth.**

Utilizing the Roman exposure calculator[1] and assuming 100-sec visits (30-sec exposure + 70-sec slew and settle), we estimate as little as four months, not necessarily contiguous, are required to cover the LSST WFD footprint to *mAB* = 25.5 in *F146* (about 1.5 mag deeper than the Euclid survey). Depending on strategic choices, the survey may take as long as 10 months, yet, **as it is not required to be done at any specific cadence or within a specific time, and images collected for this survey can serve other purposes and be collected as a part of other proposed surveys; its flexibility makes it effectively low cost in terms of Roman's observing time.** This proposal leads to a general and powerful survey that will have the potential to support many science goals proposed through the first phase of the Roman Core Community Surveys outreach[2]. Here we present a few of the science cases that would benefit from this survey, indicating optimal and acceptable strategies for each.

**Photo-z:** Photometric redshifts (photo-z) will be necessary to enable cosmology, galaxy studies, and extragalactic time-domain astronomy in the LSST era, where a wealth of photometry will be available but spectra will be comparatively sparse [11, 12, 13]. Adding NIR information to the optical 6-band LSST photometry, even in a single band, improves the characterization of all objects detected in >=1 band in each survey. While the most significant enhancements may come from a joint survey that combines all *ugrizy* and *F129-F146-F158-F213* (the y band throughput of Rubin is more effective than the corresponding *F106* Roman band) **surveying the sky in *F146*, which is the most efficient choice, would significantly improve the determination of photo-z's** (see Figure 2)**.** In particular, the addition of a Roman band to the LSST improves the photo-z estimation decreasing the fraction of outliers and the robust standard deviation moderately at low-redshift and *significantly* at intermediate and high redshifts. The *F129* filter gives a somewhat smaller photo-z scatter but is far less efficient. With the Roman exposure calculator,[3] we find that *F129* would require roughly doubling the per-image exposure time compared to *F146*, leading to a ~10-month survey.

**Deblending, crowded field photometry, and shape characterization:** Source blending is a key challenge for galaxy morphology and weak lensing (58% of the galaxies in the shallower HSC Wide survey are blended [14]). Similarly, crowded field photometry is far from a fully solved problem at LSST depths. In the extragalactic sky, blending will affect centroids, morphologies, photometry, photo-z's, and even raw source counts of galaxies. In our Galaxy, magnitude, color, astrometry, and variability inference of stars will all be impacted. Roman's PSF in optical bands is one order of magnitude smaller than LSST's and the nm-level wavefront stability of the space-based mission will allow for a very stable and well-characterized PSF. High-resolution imaging from Roman opens up joint-processing options that enable strategies to mitigate the blending effects. **Although a limited overlap of the two surveys would enable quantification of the systematic biases resulting from blending uncertainties (***e.g.* comparison of mean ensemble shears for the same set of galaxies can provide an estimate of the overall shear calibration error due to imperfect PSF corrections [3])**, the LSST-footprint-matched survey proposed here would enable quantification of the blending of *all* LSST sources (up to the magnitude limit of the shallowest of the two surveys),** and sample the full space of surface brightness and proximities in the sky. The effectiveness of a joint Rubin-Roman analysis at the pixel level in deblending sources is demonstrated in [3]: **20-30% of the objects identified in LSST images with *i*-band magnitudes brighter than 25 will be identified as multiple objects in Roman images.** Furthermore, the technical capabilities to enable this analysis are rapidly growing within the Rubin Data Management team and the scientific community at large (*e.g.* [15]). We note that white papers addressing the advantages of joint cosmological Rubin/Roman surveys include the submissions by Hirata *et al*. and Newman *et al*. Similar arguments can be made for crowded field photometry: Roman detections would benefit from the high angular resolution that would mitigate crowding confusion and would enable much improved forced-photometry measurements in LSST images.

---

[1] https://roman.gsfc.nasa.gov/science/ETC2/ExposureTimeCalc.ipynb.zip

[2] Relevant pitches with similar observing strategies as the one advocated here include booth surveys for time-domain as well as non-time-varying science. Our proposed strategy would support in full or in part (for example by providing observations in one filter in a multifilter survey) the pitches submitted by Andreoni and Fraser (producing templates for transient surveys) Li, Bedin, Dage, LeRoy, and many more.

[3] https://roman.gsfc.nasa.gov/science/ETC2/ExposureTimeCalc.ipynb.zip

**Galactic Plane Science:** Disentangling the disk and bulge components of the Milky Way has the potential to reveal the formation and evolution history of our Galaxy. Although significant progress has been achieved over the last few decades with the advent of modern astronomical surveys, observations of the central parts of the Galaxy are difficult due to substantial dust extinction (about 20 magnitudes in the *r* band towards the Galactic center) and crowding confusion for ground-based data. To date, most studies are based on exploring luminous stellar populations, such as red giants and variable RR Lyrae stars. If a similar mapping of the disk and bulge components could be done with several orders of magnitude more numerous main-sequence stars, the spatial resolution of resulting maps would be increased by at least an order of magnitude. The interstellar dust extinction decreases with wavelength: towards the Galactic center, it drops to ~8 in the LSST *y*-band and to about 5 in the Roman *F146* band (e.g., Table 1 in [16]). **A blue turn-off star in the Galactic center would be detected with SNR=5 in the coadded *y* band LSST data (assuming the same total exposure time as for the main WFD survey) and at SNR>=50 at about *F146*=22 by the Roman survey proposed here. The same data set would also yield a definitive measurement of the outer extent of the Galactic disk and map its suspected warping to exquisite detail, map disk substructure such as spiral arms and remnants of merged dwarf galaxies, and extend the limiting distance of existing 3-dimensional interstellar dust maps by a factor of several.** Milky Way science and dark matter studies would benefit from such a survey through improvements in star-galaxy separation as addressed in detail in a submission by Bechtol *et al*. As we argued for photo-z's, the scientific return of having any of the near-infrared Roman bands is strong, and the sensitivity to filter choices does not dominate the inference, and thus we again argue for the most efficient one, *F146*.

**Differential Chromatic Refraction:** Atmospheric-aided observations help constrain the colors and spectral energy distributions (SED) of galaxies (*e.g.* for AGNs [17, 18] and weak lensing [19] studies) and stars (*e.g.* flares). The atmosphere-free Roman observations, combined with its high spatial resolution and PSF characterization, would provide an accurate localization of each source observed. Accurate positional characterization for stellar and galactic sources with Roman will enable DCR characterization in the LSST data (down to the detectability limit of the Roman survey) allowing for measurements of filter-dependent displacement which in turn can be used to characterize the SED (*e.g.* for AGNs) and temperature and their time-evolution, even for rapidly evolving phenomena like stellar flares[4]. While there is some chromatic dependency on the Roman PSF, this is small and a wide filter (*F146*) does not compromise the quality of DCR measurements from the joint survey.

With the examples above, and with many more science cases that can benefit from joint analysis at the pixel or catalog level, **we advocate for a wide survey that covers the LSST WFD footprint, in a single band. This survey will enrich Rubin science by bringing in an extension of the wavelength range and increased spatial resolution, and Roman science by bringing in a time-domain dimension and shorter-wavelength photometry. This survey bootstraps on and augments the planned limited-area deep Roman surveys which already overlap parts of the LSST, but enables a full- (half-) sky legacy for Roman and extends the benefits of the joint survey results to the full southern hemisphere.**

**Suggested strategy:** The footprint of the proposed Roman survey maximizes overlap with the LSST to extend the area over which joint processing is possible and the number of objects therein that can be studied in optical and NIR: **the minimal observational strategies include *F146*=25 observations over the full ~18,000 sq deg of the WFD footprint** [9]. As we argued above, the *F146* band gives the best combination of observing efficiency and science return. **The minimum magnitude of this survey should be *F146*=25** (5-σ for point sources) to match the depth of the weak lensing gold sample and the depth of the shallowest Rubin bands (*y*) to see significant benefits in photo-z, **which would take 4-5 months**. The main limitation of this survey is the magnitude limit of the Roman data, which sets the faintest objects for which a joint analysis can be performed (note that detection is not needed in all LSST bands to perform joint analysis, but it is of course needed in at least one Roman NIR band). With a 10-month investment, the survey could choose to optimize filter selection for photo-z (which we showed prefers *F129 slightly*) or other science cases, but the extra time could also be used to increase the magnitude limit: doubling the exposure time in *F146* would get the Roman survey closer to *mAB*~26[5]. The science cases we reviewed demonstrate that going deeper has more significant advantages over optimizing the filter choice for one or a few science cases.

---

[4] Whereas Rubin will preferentially observe at low air masses to maximize image quality, a long tail of high airmass observations is expected, enabling atmosphere-aided studies of temperature and SED.

[5] Scaling under idealized conditions, using the Roman exposure calculator https://roman.gsfc.nasa.gov/science/ETC2/ExposureTimeCalc.ipynb.zip

**Table 1:** magnitude limits for LSST based on [10]

| 5-σ Depth 1 image | Median | RMS |
|---|---|---|
| All sky u band | 23.57 | 0.35 |
| All sky g band | 24.36 | 0.41 |
| All sky r band | 23.95 | 0.47 |
| All sky i band | 23.37 | 0.47 |
| All sky z band | 22.76 | 0.39 |
| All sky y band | 21.98 | 0.34 |

| 5-σ Depth COADDED | Median | Rms |
|---|---|---|
| WFD u band | 25.45 | 6.78 |
| WFD g band | 26.54 | 5.13 |
| WFD r band | 26.71 | 3.78 |
| WFD i band | 26.2 | 2.9 |
| WFD z band | 25.51 | 2.24 |
| WFD y band | 24.73 | 1.88 |

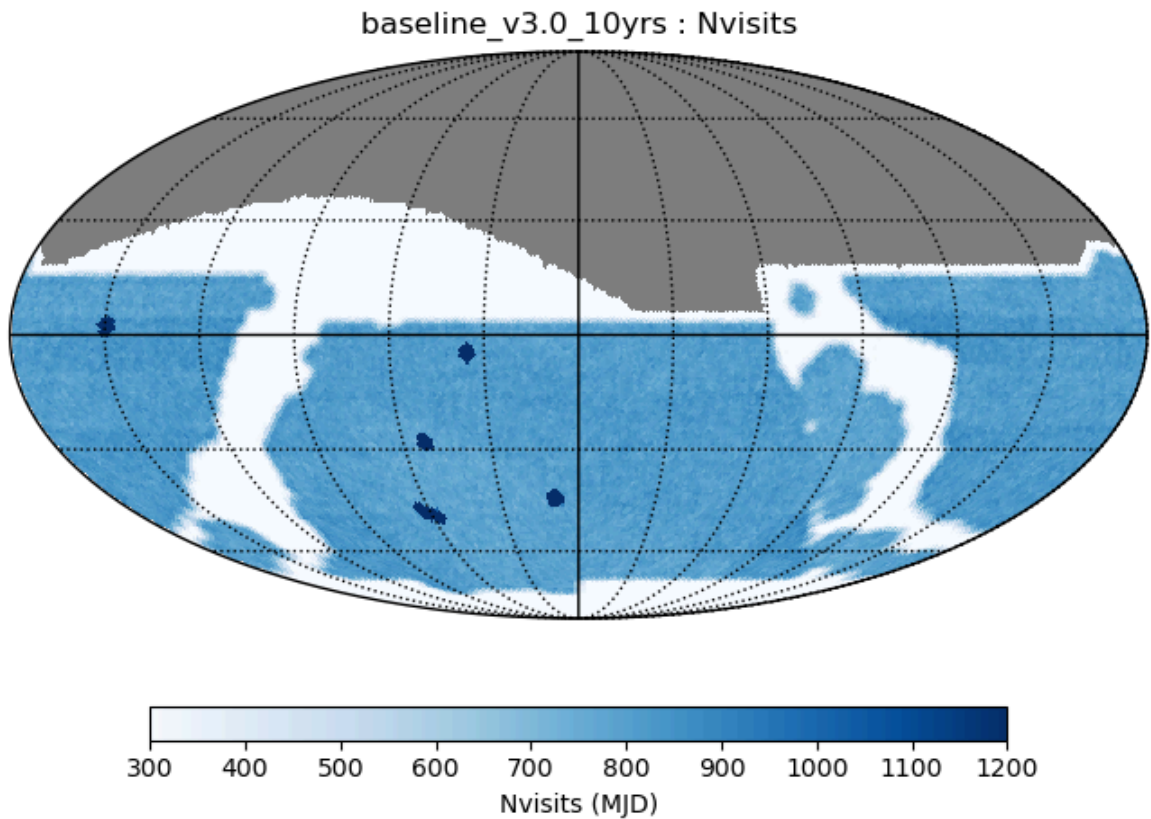

**Figure 1**: The footprint of the Rubin LSST survey as currently proposed [9]. The Wide-Fast-Deep (WFD) area of 18,620 sq. degrees receives ~800 observations over the 10 years of LSST (and appears light blue in the figure). Other special regions include the North Ecliptic Spur, South Celestial Pole (with fewer images overall, appearing in white), and Deep Drilling Fields, five selected pointings which will receive enhanced cadence observations (which appear in dark blue and receive as many as 40,000 visits).

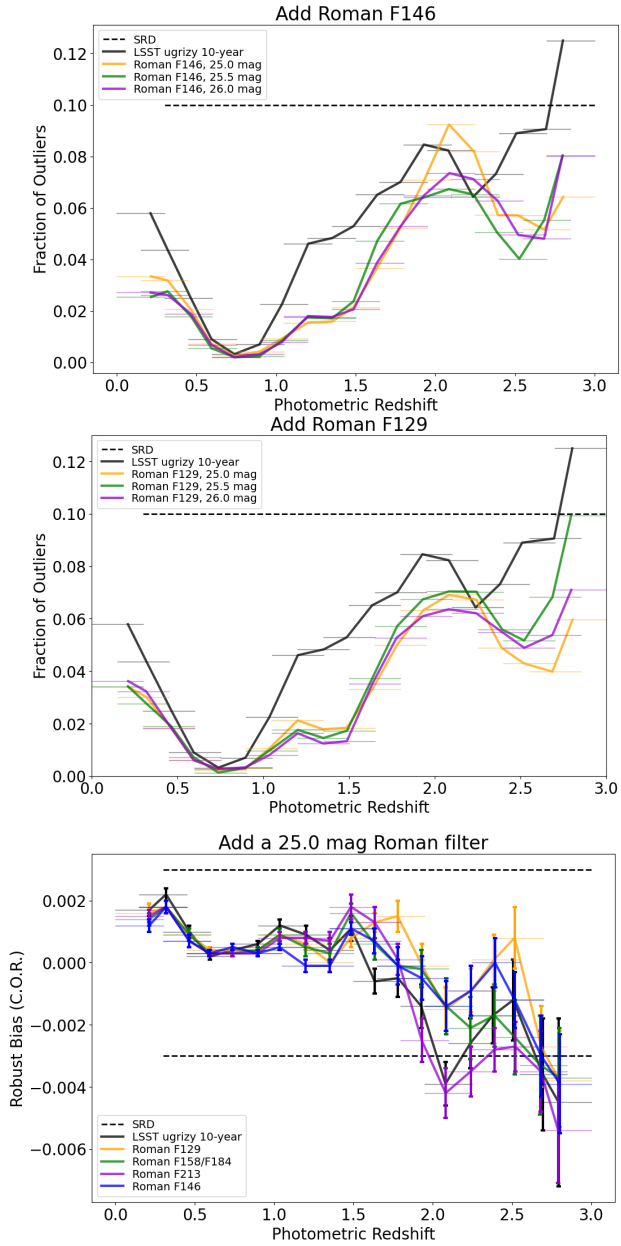
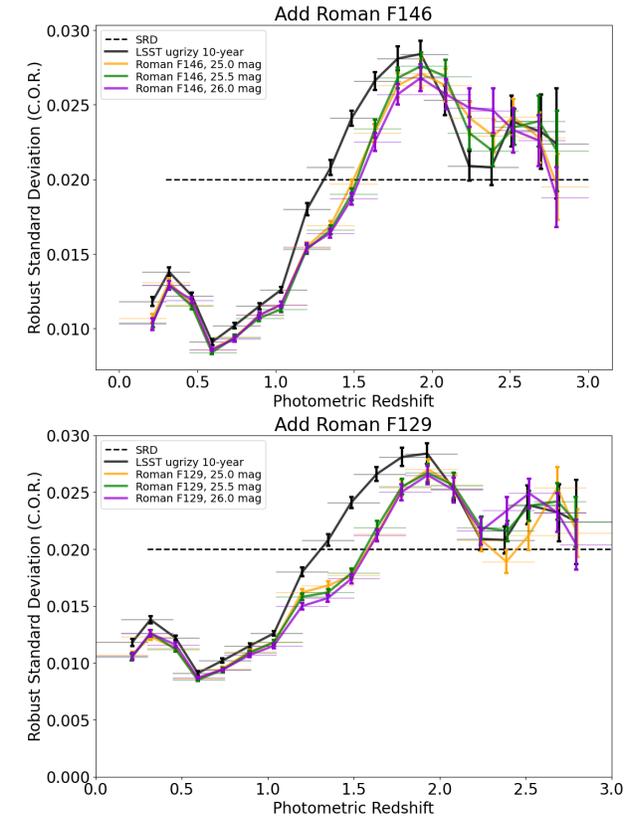

**Figure 2** Color-Matched Nearest-Neighbours (CMNN) photometric redshift (photo-z) estimator calculated as in [11, 12]: improvements in the fraction of outliers (*left*) and standard deviation (*right*) with *F146* (*top*, the proposed observing band) and *F129* filters (*middle*). The *bottom* plot shows improvements in the bias with the addition of each Roman band. Photo-z's are simulated for an LSST 10-year mock catalog (restricted to galaxies with an apparent magnitude *i*<25 mag) with 5σ limiting magnitudes of 26.1, 27.4, 27.5, 26.8, 26.1, and 24.9 in *ugrizy*, plus Roman bands (as labeled). The LSST 10-year depth and projected Roman depths are used to generate appropriate apparent magnitudes and magnitude uncertainties. Photo-z are estimated for each galaxy in turn, using the CMNN estimator with a leave-one-out approach [11]. Improvements in all metrics are notable, and generally comparable, with both *F146* and *F129* filters at a Roman 5σ limiting magnitude of 25. Deeper Roman exposures do not lead to significant improvements.